\RequirePackage{ifpdf}
\ifpdf % We are running pdfTeX in pdf mode
\documentclass[pdftex]{sigma}
\else
\documentclass{sigma}
\fi

\numberwithin{equation}{section}

\begin{document}

\allowdisplaybreaks

\renewcommand{\PaperNumber}{044}

\FirstPageHeading

\renewcommand{\thefootnote}{$\star$}

\ShortArticleName{A Discretization of the Nonholonomic Chaplygin Sphere Problem}

\ArticleName{A Discretization of the Nonholonomic\\ Chaplygin Sphere Problem\footnote{This paper
is a contribution to the Vadim Kuznetsov Memorial Issue
`Integrable Systems and Related Topics'. The full collection is
available at
\href{http://www.emis.de/journals/SIGMA/kuznetsov.html}{http://www.emis.de/journals/SIGMA/kuznetsov.html}}}

\Author{Yuri N. FEDOROV}

\AuthorNameForHeading{Yu.N. Fedorov}

\Address{Department de Matem\'atica I, Universitat Politecnica de Catalunya, Barcelona, E-08028, Spain}
\Email{\href{mailto:Yuri.Fedorov@upc.edu}{Yuri.Fedorov@upc.edu}}
\URLaddress{\url{http://www.ma1.upc.edu/personal/fedorov.html}}

\ArticleDates{Received December 13, 2006, in f\/inal form February
26, 2007; Published online March 12, 2007}

\Abstract{The celebrated problem of a non-homogeneous sphere
rolling over a horizontal plane was proved to be integrable and
was reduced to quadratures by Chaplygin. % in \cite{Chapl}.
Applying the formalism of variational integrators (discrete
Lagrangian systems) with nonholonomic constraints
and introducing suitable discrete constraints, we construct a
discretization of the $n$-dimensional generalization of the Chaplygin
sphere problem, which preserves the same f\/irst integrals as the
continuous model, except the energy.
We then study the discretization of the classical 3-dimensional problem for a class of special
initial conditions, when an analog of the energy integral does exist and the corresponding map
is given by an addition law on elliptic curves. The existence
of the invariant measure in this case is also discussed.}


\Classification{37J60; 37J35; 70H45}

\begin{flushright}
{\it Dedicated to the memory of Vadim, who made \\  me fascinated with discrete integrable systems}
\end{flushright}

\renewcommand{\thefootnote}{\arabic{footnote}}
\setcounter{footnote}{0}

\section{Introduction} The Chaplygin problem on a non-homogeneous sphere rolling over a
horizontal plane without slipping is probably one of the best
known integrable systems of the classical nonholonomic mechanics.
Although being non-Hamiltonian in the whole phase space (see \cite{BM2}),
the equations of motion possess an invariant
measure, which is a rather strong property putting them rather
close to Hamiltonian systems.

It is even more interesting that the Chaplygin sphere appears to
be an {\it algebraic integrable} system in the sense that generic
level varieties of the f\/irst integrals are open subsets of
2-di\-men\-sional Abelian varieties and, {\it after an appropriate
time reparameterization}, the phase f\/low becomes a straight line
uniform on them, \cite{Dust, Chapl_Fed}.

Note that a Lax pair with a spectral parameter for the Chaplygin sphere,
which could provide all the constants of motion,
is still unknown and, probably does not exist. Hence, one cannot use the powerful method of
Baker--Akhieser functions to f\/ind theta-function solution of the problem or to construct its integrable
discretization by applying a B\"acklund transformation, as described in \cite{KV,JNMP_Fed},
or by factorization of matrix polynomials.

{\bf Contents of the paper.} We brief\/ly recall the equations of motion of the Chaplygin
sphere, as well as their $n$-dimensional generalization. Then we
construct a discretization of the problem by applying the formalism of variational integrators
with nonholonomic constraints recently developed in \cite{CM1,deLeon1}.
Apart from the nonholonomic distribution on the tangent bundle
$T\,Q$ of the conf\/iguration manifold $Q$, the formalism requires
introducing discrete constraints on $Q\times Q$ which, in a
certain sense, must be consistent with the continuous ones. As one can
see in the li\-te\-ra\-ture \cite{CM1, FeZe2004, Perl}, the choice
of discrete constraints is crucial for the behavior of the
discretized nonholonomic system: the latter may inherit the main
properties of its continuous counterpart, or may not. As an
example, we note that although continuous systems with stationary
nonholonomic constraints possess the energy integral, almost all
their known discretizations do not enjoy this property (see e.g.,
\cite{Cort2}).
Nevertheless, for a class of discretizations considered in \cite{FeZe2004} there exists a natural method to
specify discrete constraints which ensures the exact preservation of energy.

Our choice of discrete nonholonomic constraints on $E(n)\times E(n)$
that mimics the condition of the sphere rolling without slipping over a horizontal plane
allows us to construct a map
discretizating the Chaplygin sphere problem, which has the form of a momentum conservation law and therefore
preserves all the momenta integrals of the original system.

We then consider the discretization of the classical 3-dimensional problem for the case when the discrete
angular momentum is vertical. In this special case the structure of the map is reminiscent
to that of the Veselov--Moser discretization of the Euler top on $SO(3)$ \cite{Ves, MosVes}
and an analog of the quadratic energy integral does exist.

This implies that the invariant manifolds of the discretization
map are elliptic curves $\mathcal E$ and the map is described as
an addition law. However, in contrast to most known integrable
discrete systems, in the discrete Chaplygin sphere the translation
on $\mathcal E$ depends not only on the constants of motion but
also on the point on the curve. Thus, in order to f\/ind the
explicit solution, we arrive at a rather dif\/f\/icult problem of
reparameterization of $\mathcal E$ or its real part, which would
make the translation constant. We notice that this problem is
equivalent to the problem of the existence of an invariant measure
of the map.

\section{The Chaplygin sphere and its multidimensional generalization}
Following Chaplygin \cite{Chapl}, consider a dynamically
non-symmetric sphere with inertia tensor $J$, radius $\rho$, and
mass $m$ rolling without slipping over a horizontal plane. Assume
that the mass center and the geometric center $C$ of the sphere
coincide. Let $\gamma=(\gamma_1,\gamma_2, \gamma_3)^T$ be the
vertical unit vector and $\omega=(\omega_1, \omega_2, \omega_3)^T$,
$v\in {\mathbb R}^3$ be respectively the angular velocity of the
sphere and the velocity of its center in the moving frame. The
condition of non-slipping of the point $P$ of contact of the
sphere with the horizontal plane is
\begin{gather}
\label{vel}
v+\rho \omega\times\gamma =0 .
\end{gather}
Here and below, $\times$ denotes the vector product in ${\mathbb R}^3$.
On the conf\/iguration space of the problem, the Lie group $E(3)$, these equations def\/ine {\it nonholonomic} constraints:
for any two positions of the sphere on the plane there exists a linking trajectory that satisf\/ies \eqref{vel}.

Under these constraints the equations of motion can be reduced to the following closed system for $\omega$, $\gamma$
\cite{Chapl,FeKo2}\footnote{We use our proper notation
for the variables and parameters of the problem, which may dif\/fer from that used in the
original paper \cite{Chapl}.}
\begin{gather}
\Lambda\dot\omega =\Lambda\omega\times\omega+
\frac D F \,\langle \Lambda^{-1}\gamma, (\Lambda\omega)\times\omega\rangle \gamma,%\nonumber
\qquad
\dot\gamma =\gamma\times\omega,
\label{nh4.7}
\end{gather}
where
\begin{gather*}
D=m \rho^2, \qquad \Lambda=J+D{\mathbb I}, \qquad F= \langle J\gamma,\Lambda^{-1}\gamma\rangle
\equiv 1- D\langle \gamma,\Lambda^{-1}\gamma\rangle ,
\end{gather*}
$\mathbb I$ being the identity $3\times 3$ matrix.
Let $K=\Lambda\omega-D\langle \omega,\gamma\rangle \gamma$ be the vector of the
angular momentum of the sphere with respect to the contact point $P$.
Then the system \eqref{nh4.7} also admits representation in form
\begin{gather}
\label{vectors}
\dot K=K\times \omega, \qquad \dot \gamma=\gamma \times \omega .
\end{gather}
Hence, like $\gamma$, the momentum $K$ is f\/ixed in space. As a
result, the system possesses four independent f\/irst integrals
\begin{gather}
\label{int0}
\langle \gamma,\gamma\rangle =1, \qquad \langle K,\gamma\rangle
\equiv \langle J\omega,\gamma\rangle =h, \qquad \langle K,K\rangle =n, \qquad
\langle K,\omega\rangle =l .
\end{gather}
Here the last integral is the kinetic energy of the sphere  and, since
\begin{gather*}
%\label{oms1}
\Lambda \omega= K - \frac {D}{F} \left\langle \gamma, \Lambda^{-1} K\right\rangle \gamma,
\end{gather*}
it can be rewritten in terms of $K$, $\gamma$:
\begin{gather}
\label{en_int}
\langle K, \Lambda^{-1} K\rangle -\frac DF \langle \gamma, \Lambda^{-1} K \rangle ^2 .
\end{gather}

In addition, the system \eqref{nh4.7} possesses an invariant
measure
\begin{gather} \label{inv_m}
\sqrt{\langle J \gamma,\Lambda^{-1}\gamma\rangle }\, d\omega_1 \wedge \cdots \wedge d\gamma_3,
\end{gather}
hence, by the Euler--Jacobi theorem (which is also often refereed
to as the Jacobi last multiplier theorem, see e.g., \cite{Whitt}),
it is integrable by quadratures and its generic invariant
varieties are 2-dimensional tori.

There are two special types of the initial conditions, when the
equations of motion are simp\-lif\/ied. In the f\/irst case the momentum $K$
is horizontal, $\langle K, \gamma\rangle =0$. As shown by
Chaplygin, after a time reparameterization and introducing
spheroconic coordinates on $S^2=\langle\gamma,\gamma\rangle$ the
variables separate and the system reduces to hyperelliptic
quadratures. Theta function solution for the unreduced Chaplygin
system in this, as well as in the generic case, was obtained in
\cite{Chapl_Fed}.

The second special case is described below.

{\bf The case of the vertical momentum $\mathbf{\emph{K}}$.}
As noticed in \cite{Chapl} and as follows from
the energy integral in \eqref{int0}, under the special initial conditions $K=h\gamma$,
$h={\rm const}$ one has $ \langle K, \omega \rangle= h\langle \gamma, \omega \rangle=l$, i.e.,
\begin{gather*}
%\label{fix}
\langle \omega,\gamma\rangle = \frac {l}{h}=\mbox{const},
\qquad \mbox{and} \qquad \Lambda \omega=\frac{h^2+Dl}{h} \gamma.
\end{gather*}
As a result, the f\/irst vector equation in \eqref{nh4.7} takes the form
of the Euler top equations % for the momentum $M=\Lambda \omega$,
\begin{gather}
\label{euler}
\Lambda\dot \omega=\Lambda \omega\times \omega ,
\end{gather}
and integrals \eqref{int0} reduce to
\begin{gather*}
%\label{Eu_int}
\langle \omega, \Lambda\omega\rangle =l, \qquad \langle \Lambda\omega, \Lambda\omega\rangle =k^2, \qquad
l,k=\mbox{const}.
\end{gather*}
Hence, for almost all initial conditions $K=h\gamma$, the variables
$\omega_i$, $\gamma_i$ are elliptic functions of the original time $t$ and the
solution of the reduced system is periodic.

{\bf Multidimensional generalization.}
Now, following \cite{FeKo2},  consider the generalized Chaplygin problem on an
$n$-dimensional ball rolling without slipping on an $(n-1)$-dimensional `horizontal'
hyperplane $H$ in ${\mathbb R}^{n}$. The conf\/iguration space for the ball is the Lie group
$E(n)=(R, r)$, where $R\in SO(n)$, $r\in {\mathbb R}^n$ are the rotation matrix of the sphere
and the position vector of its center~$C$. For a trajectory $R(t)$, $r(t)$, def\/ine the Lie algebra
element $({\mathbf{\omega} }, v_C)$, where
\begin{gather}
\label{Lie}
{\mathbf{\omega} }= R^{-1} \dot R \in so(n), \qquad v_{C}=R^{-1} \dot r \in {\mathbb R}^{n}
\end{gather}
are respectively the angular velocity and the velocity of $C$ in the frame attached to the sphere.

Let now $\gamma\in{\mathbb R}^{n}$ be the unit vector orthogonal to the hyperplane
$H$ and directed `upwards', i.e., from $H$ to the center $C$, and, as above,
$\rho$ be radius of the ball. Then the condition for the sphere rolling without slipping that generalizes
\eqref{vel} reads
\begin{gather}
\label{cont_const}
v_{C}+ \rho \, {\mathbf{\omega} } \, \gamma =0 .
\end{gather}
One can show that this vector constraint determines a non-integrable distribution on the tangent bundle
$T\, E(n)$, which is neither left- no right-invariant with respect to the action of $E(n)$.

Next, introduce the angular momentum of the sphere {\it in the body frame} with respect to the center $C$,
\begin{gather}
\label{cont_mom}
M=J{\mathbf{\omega} } +{\mathbf{\omega} } J \in so^{*}(n),
\end{gather}
$J$ being a constant diagonal $n\times n$ matrix. Then, as shown in \cite{FeKo2}, the motion of the sphere
is described by the following Euler--Lagrange equations
\begin{gather}
\dot M+[{\mathbf{\omega} },M]=\rho F \wedge \gamma  ,\qquad m(\dot v_{C}+{\mathbf{\omega} } \, v_{C})=F, \qquad
\dot\gamma+{\mathbf{\omega} } \gamma =0 ,
\label{md2.15}
\end{gather}
where $F\in {\mathbb R}^{n}$ is the reaction force acting at the point $P$ of contact of the sphere with the
hyperplane $H$. Here all the vectors are considered in the frame attached to the ball.

Now, dif\/ferentiating the constraints \eqref{cont_const}
and using the second equation in \eqref{md2.15}, we f\/ind
$F=m\rho {\mathbf{\dot \omega} }\gamma$. Then \eqref{md2.15} gives
\begin{gather*}
\dot M +[{\mathbf{\omega} },M]=D( {\mathbf{\dot \omega} } \Gamma+ \Gamma { \mathbf{\dot\omega} }), \qquad
\dot \Gamma+[{\mathbf{\omega} } ,\Gamma ]=0 ,
\end{gather*}
where $D=m\rho^{2}$, $\Gamma= \gamma \otimes \gamma \equiv \gamma \gamma^T$.
Then this system can be represented in the
following compact commutative form that generalizes \eqref{vectors}
\begin{gather}
\label{mat}
\dot {\bf {K}}+[{\mathbf{\omega} },{\bf {K}}]=0 ,\qquad  \dot\Gamma +[{\mathbf{\omega} } ,\Gamma ]=0 ,
\end{gather}
where
\begin{gather}
\label{K}
{\bf {K}}=J{\mathbf{\omega} } +{\mathbf{\omega} } J+D(\Gamma {\mathbf{\omega} } +{\mathbf{\omega} } \Gamma )\in so^*(m)
\end{gather}
can be regarded as
the angular momentum of the ball relative to the contact point $P$.

It is natural to introduce the nondegenerate inertia operator
${\mathcal A}\colon so(n)\rightarrow so(n)^{*}$ such that
${\bf {K}}={\mathcal A}{\mathbf{\omega} }$.
Since $\mathcal A$ is nondegenerate, equations \eqref{mat} give a
closed system for the variables~${\mathbf{\omega} }$,~$\gamma$.

As follows from the form of \eqref{mat},
${\rm Ad}_{R}{\bf {K}}$ is a constant tensor in the space.
Hence this system has the following set of integrals
\begin{gather}
{\rm tr}\; {\bf {K}}^{s}, \qquad {\rm tr}\; ({\bf {K}}^{s}\Gamma ^{l}) ,\qquad
{\rm tr} \; \Gamma =1, \qquad  s=2,4,6,\ldots, \qquad l\in{\mathbb N} .
\label{md2.19}
\end{gather}
naturally, the system also possesses the energy integral
\begin{gather}
H = -\frac 14 {\rm tr} ({\mathbf{\omega} } {\cal A}{\mathbf{\omega} } )
\label{md2.20}
\end{gather}
and, as shown in \cite{FeKo2},
the invariant measure $\mu\, d{\mathbf{\omega} } \wedge d\gamma$ with density $\mu=\sqrt{\det {\mathcal A}}$,
which depends on the components of $\gamma$ only.

In the classical case $n=3$, under the isomorphism between $so(3)$
and ${\mathbb R}^3$,
\begin{gather*}
{\mathbf{\omega} }_{ij}=\varepsilon_{ijk} \omega_k , \qquad  {\bf {K}}_{ij}=\varepsilon_{ijk} K_k
\end{gather*}
the integrals \eqref{md2.19}, \eqref{md2.20} and the measure transform to \eqref{int0} and \eqref{inv_m}
respectively.

Various properties of the Chaplygin sphere and its $n$-dimensional generalization were studied in
\cite{Dust, Kilin, Chapl_Fed, Schneider}, although
the integrability or nonintegrability of this generalization was not proved.

\section{Discrete mechanical systems with nonholonomic constraints}

Before describing the discrete setting, let us recall that
a continuous nonholonomic Lagrangian system is a triple $(Q,L,{\mathcal D} )$,
where $Q$ is a smooth $n$-dimensional conf\/iguration space, $ L:TQ \to \mathbb R $ is a smooth function
called the {\em Lagrangian}, and ${\mathcal D} \subset TQ$ is a $k$-dimensional
{\em constraint distribution}. Let $q=(q_1 ,\dots, q_n)$ be local
coordinates on $Q$. In the induced coordinates $ (q, \dot q)$ on the
tangent bundle $TQ$ we write $ L(q ,\dot q) $. It is assumed that
the map
\begin{gather*}
\frac{ \partial  L}{\partial \dot q }\colon TQ \to T^*Q
\end{gather*}
is invertible.
A curve $q(t) \in Q$ is said to {\em satisfy the
constraints} if $\dot{q}(t)\in{\mathcal D}_{q(t)}$ for all~$t$.
The equations of motion are
given by the La\-gran\-ge--d'Al\-em\-bert principle:
\begin{gather}
\label{LD.eqn}
\delta \int^b_a L(q ^i , \dot{q} ^i ) \, d t = 0,
\end{gather}
where $\delta q(t) \in {\mathcal
  D}_{q(t)}$ for $t\in (a,b)$ and $\delta q(a) = \delta q(b) = 0$.
This principle is
supplemented by the condition that the curve itself satisf\/ies the
constraints. Note that we take the variation {\em before} imposing the
constraints.

Assuming that the constraint distribution is specif\/ied by a set of
dif\/ferential forms $ A^j(q)$, $j = 1, \dots, s<n$,
\begin{gather}
\label{00}
{\mathcal D} = \{\dot q \in TQ \mid \langle A ^j(q), \dot q \rangle =
0,\ j=1,\dots,s \},
\end{gather}
the principle \eqref{LD.eqn} implies the Euler--Lagrange equations with multipliers $\lambda_j$
\begin{gather}
\frac{d}{dt}\frac{\partial L}{\partial \dot q} - \frac{\partial L}{\partial q}
= \sum_{j=1}^s \lambda_j A ^j(q).
\label{1.1}
\end{gather}
Coupled with \eqref{00}, they give a complete description of the dynamics of the system.

Note that equations \eqref{1.1} conserve the energy
\begin{gather*}
%\label{energy.eqn}
E = \biggl\langle \frac{\partial L}{\partial \dot q} , \dot q \biggr\rangle - L.
\end{gather*}

{\bf The discrete Lagrange--d'Alembert principle and equations.}
Let, as above, $Q$ be a~smooth manifold. According to \cite{CM1},
a~discrete nonholonomic mechanical system on $Q$ is def\/ined by
three ingredients
\begin{itemize}\itemsep=0pt
\item[(1)]
a {\em discrete Lagrangian} ${\mathbb L}\colon Q\times Q \to {\mathbb R}$;

\item[(2)]
an $(n-s)$-dimensional distribution $\cal D$ on $TQ$ (given by equations \eqref{00});

\item[(3)] a discrete constraint manifold ${\mathcal D}_d\subset
  Q\times Q$, which has the same dimension as $\cal D$ and satisf\/ies the
  condition $(q,q) \in \mathcal D_d$ for all $q \in Q$.
\end{itemize}
The dynamics is given by the following {\em discrete
  Lagrange--d'Alembert principle} (see \cite{CM1}),
\begin{gather*}
\sum _{k = 0} ^{N-1}
(D _1 {\mathbb L} ( q _k , q _{k+1}) + D _2 {\mathbb L} (q _{k-1} , q _k))\,
 \delta q_k = 0, \qquad \delta q _k \in {\mathcal D} _{q _k }, \qquad
(q_k , q_{k+1}) \in {\mathcal D}_d.
\end{gather*}
Here $D_1 {\mathbb L} $ and $ D_2 {\mathbb L} $ denote the partial derivatives of
the discrete Lagrangian with respect to the f\/irst and the second
inputs, respectively.

The discrete constraint manifold is specif\/ied by the discrete constraint functions
\begin{gather}
\label{gen.constr}
{\cal F}_j(q_k, q_{k+1})=0, \qquad j=1,\dots, s,
\end{gather}
which impose the above restriction on the solution sequence $\{(q_k, q _{k+1} )\}$.

\begin{remark}
If the discrete Lagrangian $\mathbb L$ is obtained from a continuous one,
$L(q,\dot q)$, via a~discretization mapping $\Psi\colon Q\times Q\to TQ$
def\/ined in a neighborhood of the diagonal of $Q\times Q$, i.e.,
${\mathbb L}=L\circ \Psi$, then the variety ${\mathcal D}_d$ must be {\it
  consistent} with the continuous distribution ${\mathcal D}$. Namely,
${\mathcal D}_d$ is locally def\/ined by the equations $A^j \circ \Psi = 0$, $ j= 1, \dots, s$.

In particular, if the distribution ${\mathcal D}\subset T\,Q$ is integrable, i.e., it def\/ines an $(n-s)$-dimensional
integral submanifold ${\cal N}\subset Q$, then ${\mathcal D}_d\subset Q\times Q$ is just the product
${\cal N}\times{\cal N}$.
\end{remark}

We emphasize that, in general, the
discretization mapping is not unique and hence there are many ways to
def\/ine the discrete Lagrangian $ {\mathbb L} $ and the discrete constraint
manifold ${\mathcal D} _d $ for a~given nonholonomic system $(Q, L,
{\mathcal D})$.\footnote{An alternative approach to the discretization
  of nonholonomic systems based on a modif\/ication of canonical
  transformations was proposed in \cite{deLeon1}.}

The discrete principle implies that the solutions of a discrete nonholonomic system are represented by
sequences $ \{(q _k, q _{k+1})\}$ that satisfy the
  discrete Lagrange--d'Alembert equations with multipliers
\begin{gather}
\label{dis_LA.eqn}
D_1 {\mathbb L} (q_k, q_{k+1})+ D_2 {\mathbb L} (q_{k-1},q_k)
= \sum_{j=1}^s \lambda_{j}^k A_j (q_k) .
% {\mathcal F}_j(q_k, q_{k+1}) =0.
\end{gather}
The multipliers $\lambda_{j}^k$ are determined from the discrete constraints \eqref{gen.constr},
but, in general, not uniquely. Hence, the map $Q\times Q \mapsto Q\times Q$ def\/ined by
\eqref{gen.constr}, \eqref{dis_LA.eqn} is multi-valued.

\begin{remark}
According to \cite{CM1}, equations \eqref{dis_LA.eqn} introduce a
well-def\/ined mapping %$\Phi: Q\times Q \to Q\times Q$,
$(q_{k-1}, q_{k})\mapsto (q_{k}, q_{k+1})$,
if the $(n+s)\times (n+s)$ matrix
\begin{gather*} \begin{pmatrix}
D_1 D_2 L _d(q_k ,q_{k +1}) & A^1(q_k) & \cdots & A^s(q_k)  \\
D_2 {\cal F}_1 (q_k ,q_{k +1}) & 0  & \cdots & 0 \\
\vdots & \vdots &  & \vdots \\
D_2 {\cal F}_s (q_k ,q_{k +1}) & 0  & \cdots & 0
\end{pmatrix}
\end{gather*} is invertible for each $(q_k ,q_{k +1})$ in a neighborhood of the
diagonal of $Q\times Q$.
\end{remark}

\section{A discretization of the Chaplygin sphere}
Now we apply the above approach to discretize the generalized Chaplygin sphere problem on~$E(n)$. The trajectory
of such system is a sequence $(R_k, r_k)$, $k\in {\mathbb Z}$. We choose the discrete Lagrangian in form
\begin{gather}
\label{disc_L}
{\mathbb L}= \frac 12 \mbox{Tr} (R_k J R_{k+1}^T) + \frac m 2 \langle \Delta r_k, \Delta r_k \rangle,
\qquad  \Delta r_k = r_{k+1}-r_k .
\end{gather}
Here the rotational part $\frac 12 \mbox{Tr} (R_k J R_{k+1}^T)$
coincides with the Lagrangian of the Euler top on $SO(n)$
introduced in \cite{Ves}.

Note that, in view of \eqref{Lie}, the continuous constraints \eqref{cont_const} can be rewritten as
\begin{gather}
\label{cc}
\dot r + \rho \dot R R^T \vec \gamma =0 ,
\end{gather}
where $\vec \gamma $ is the unit normal vector in the f\/ixed frame
(without loss of generality we can set it to be $(0,\dots,
0,1)^T$.)

Then, in view of \eqref{disc_L}, \eqref{cc}, in our case the discrete Lagrange--d'Alembert equations
\eqref{dis_LA.eqn} take the form
\begin{gather}
R_{k+1} J+ R_{k-1} J = \Lambda_k R_k +  \rho \vec f^k \, \vec \gamma^T  R_k,
\label{Lagr_R} \\
m (r_{k+1 }  -r_k ) - m (r_k -r_{k-1}) = \vec f^k ,
\label{lin1}
\end{gather}
where $\Lambda_k$ is the symmetric matrix Lagrange multiplier and $\vec f^k=(f_1^k, \dots, f^k_n)^T$, is
the vector multiplier corresponding to the constraints \eqref{cc}.

In view of the property $R_k R_k^T ={\bf I}$, equations \eqref{Lagr_R} yield
\begin{gather}
 R_{k+1} J R_k^T + R_{k-1}J R_k^T  =\Lambda_k + \rho  \vec f^k  \vec \gamma^T,\nonumber
 \\
 R_k J R_{k+1}^T + R_k J R_{k-1}^T =\Lambda_k^T + \rho \vec\gamma \vec f^T_k .
\label{**}
\end{gather}
Following \cite{Ves}, we introduce the discrete angular momentum of the sphere with respect to its center $C$
{\it in space}
\begin{gather*}
m_k= R_{k+1} J R_k^T- R_k J R_{k+1}^T \in so(n).
\end{gather*}
Then \eqref{**}, \eqref{lin1} give rise to
\begin{gather}
\label{de2}
m_k = m_{k-1} + \rho (\vec f^k \vec\gamma^T - \vec\gamma \vec {f^k}^T), \qquad
m ( \Delta r_{k } -\Delta r_{k-1})= \vec f^k  .
\end{gather}
Next, like in \cite{Ves}, introduce the discrete momentum of the sphere in the body frame
\begin{gather*}
%\label{*}
M_k = {\rm Ad_{R_k}} m_k\equiv \Omega_k J - J\Omega_k^T , \qquad \Omega_k= R_{k}^T R_{k+1}\in SO(n),
\end{gather*}
where $\Omega_k$ is the f\/inite rotation in the body frame.
In the continuous limit $\Omega_k \mapsto {\bf I}+ \varepsilon \omega$, $\omega\in so(n)$, it transforms
to the momentum \eqref{cont_mom}.

Then the system \eqref{de2} gives
\begin{gather*}
M_k = \Omega_{k-1}^T M_{k-1} \Omega_{k-1}
+ \rho (R_k^T\vec f^k \gamma^T R_k - R_k^T\gamma_k \vec f^T R_k ) ,
\\
m ( \Delta r_{k } -\Delta r_{k-1})= \vec f^k \, ,
\end{gather*}
that is,
\begin{gather}
\label{EP1}
M_k = \Omega_{k-1}^T M_{k-1} \Omega_{k-1} + \rho \, \vec F^k \wedge \gamma^T_k, \qquad
m ( \Delta r_{k } -\Delta r_{k-1})= \vec f^k \, ,
\end{gather}
where
\begin{gather}
\label{Mult}
\vec F^k= R_k^T\vec f^k, \qquad \gamma_k =R_k^T \vec\gamma
\end{gather}
represent the vectors $\vec f^k$, $\vec\gamma$ in the body frame.
Equations \eqref{EP1} can be regarded as a discrete version of
the equations of motion of the Chaplygin sphere \eqref{md2.15}.

{\bf Discrete constraints.} In order to determine the vector multiplier $\vec f^k$
we must specify discrete constraints
on $E(n)\times E(n)$. A naive choice that imitates the form of the
continuous constraints~\eqref{cont_const} is
\begin{gather*}
\Delta r_k  + \rho \, \bar\Omega_k \bar\gamma=0,
\end{gather*}
where $\bar \Omega_k = R_{k+1} R_k^T \in SO(n)$ is the f\/inite rotation of the sphere in the {\it space frame}.
However, this choice allows that the displacement $\Delta r_k$ of the center $C$ may be not ``horizontal'' (i.e.,
orthogonal to the vertical vector $\bar\gamma$), which is incompatible with the mechanical setting of the problem.

For this reason, our choice of the corresponding discrete
constraints \eqref{gen.constr} will be
\begin{gather}
\label{dc}
\Delta r_k  + \frac  \rho 2  (\bar\Omega_k - \bar\Omega_k^T ) \vec \gamma =0 ,
\end{gather}
which does ensure that $\Delta r_k$ is orthogonal to $\bar\gamma$.
In the continuous limit $\bar\Omega_k \mapsto {\bf I}+ \omega$ it
transforms to the constraint \eqref{cc}.

Then, in view of \eqref{dc} and the second equation in \eqref{EP1},
\begin{gather*}
\vec f^k  = - \frac {m\rho}2 (\bar\Omega_k - \bar\Omega_k^T ) \vec \gamma
+ \frac {m\rho}2 (\bar\Omega_{k-1} - \bar\Omega_{k-1}^T ) \vec \gamma , \\
\vec F^k  =R_k^T\vec f^k= - \frac {m\rho}2 (\Omega_k-\Omega^T_k )\gamma_k
-  \frac {m\rho}  2 (\Omega_{k-1}-\Omega^T_{k-1} )\gamma_k ,
\end{gather*}
and, therefore,
\begin{gather*}
\vec F^k \wedge \gamma^T_k  =  - \frac {m\rho}  2 [ \Omega_k  \Gamma_k+ \Gamma_k  \Omega_k  ]
+ \frac {m\rho}  2 [ \Omega_k^T  \Gamma_k+ \Gamma_k  \Omega_k^T  ]
+ \frac {m\rho}  2 [ \Omega_{k-1}  \Gamma_k+ \Gamma_{k}  \Omega_{k-1}  ]
\\ \phantom{\vec F^k \wedge \gamma^T_k=}{}
{}- \frac {m\rho}  2 [ \Omega_{k-1}^T  \Gamma_k+ \Gamma_k  \Omega_{k-1}^T  ] ,
\end{gather*}
where, as in Section 2, $\Gamma_k= \gamma_k \otimes \gamma_k \equiv \gamma_k \gamma_k^T$.

As a result, we eliminate the multipliers $\vec F^k$ and from \eqref{EP1}, \eqref{Mult} get the equations
\begin{gather}
M_k + \frac {m\rho^2} 2 [ \Omega_k  \Gamma_k+ \Gamma_k  \Omega_k  ]
- \frac {m\rho^2} 2 [ \Omega_k^T  \Gamma_k+ \Gamma_k  \Omega_k^T  ]  = \Omega_{k-1}^T M_{k-1} \Omega_{k-1} \nonumber
\\ \phantom{M_k}
{} + \frac {m\rho^2} 2 [ \Omega_{k-1}  \Gamma_k+ \Gamma_{k}  \Omega_{k-1}  ]-
\frac {m\rho^2} 2 [ \Omega_{k-1}^T  \Gamma_k+ \Gamma_k  \Omega_{k-1}^T  ] ,
\label{EP2}
\\
\gamma_k = \Omega_{k-1}^T \gamma_{k-1} ,
\label{gam}
\end{gather}
which def\/ine an implicit map  ${\mathcal C}\colon (\Omega_{k-1},\gamma_{k-1}) \mapsto (\Omega_k, \gamma_k)$.

\begin{proposition} \label{main1} The map $\mathcal C$ admits the following compact representation
\begin{gather}
\label{EP3}
{\mathcal K}_k = \Omega_{k-1}^T {\mathcal K}_{k-1} \Omega_{k-1} , \qquad
\Gamma_k = \Omega_{k-1}^T \Gamma_{k-1} \Omega_{k-1} ,
\end{gather}
where ${\mathcal K}_k$ is the discrete analog of the momentum with respect to the contact point $P$ of the sphere,
\begin{gather}
{\mathcal K}_k  = \Omega_k \left(J+ \frac D2 \Gamma_k\right) -\left(J+ \frac D2 \Gamma_k\right) \Omega_k^T
+ \frac D2 ( \Gamma_k\Omega_k  - \Omega_k^T\Gamma_k )  \nonumber
\\ \phantom{{\mathcal K}_k}
{}\equiv M_k+ \frac D2 (\Omega_k \Gamma_k - \Gamma_k \Omega_k^T)+
\frac D2 ( \Gamma_k\Omega_k  - \Omega_k^T\Gamma_k ) , \qquad D={m\rho^2} .
\label{dis_K}
\end{gather}
\end{proposition}

Indeed, in the continuous limit $\Omega_k \mapsto {\bf I}+
\varepsilon \omega$, $\varepsilon \ll 1$ the matrix ${\mathcal K}/\varepsilon$ tends
precisely to the angular momentum \eqref{K}
of the $n$-dimensional sphere with respect to its contact point,
whereas the relations \eqref{EP3} transform to the continuous
equations \eqref{mat}.

The map $\mathcal C$ therefore is evaluated as follows:
\begin{description}\itemsep=0pt
\item{1)} given $\Omega_{k-1}$, $\Gamma_{k-1}$, determine ${\mathcal K}_{k-1}$ by \eqref{dis_K};
\item{2)} Calculate ${\mathcal K}_{k}$, $\Gamma_{k}$ by \eqref{EP3};
\item{3)} Given ${\mathcal K}_{k}$, $\Gamma_{k}$, reconstruct $\Omega_k$ as a solution of the matrix equation \eqref{dis_K}.
\end{description}

The latter equation, in general, has several solutions, hence the map $\mathcal C$ is multi-valued.

An an immediate corollary of Proposition \ref{main1}, we obtain

\begin{proposition}
Regardless to a branch of the map $\cal C$, it preserves the set of momenta integ\-rals~\eqref{md2.19} of the continuous
Chaplygin sphere problem with $\bf {K}$ replaced by $\mathcal K$.
\end{proposition}

This property gives a solid justif\/ication of our choice of the discrete constraint \eqref{dc}.
Note however that a discrete analog of the energy integral \eqref{md2.20} in the generic case
is unknown (see also Proposition \ref{main2} below).
\medskip

\noindent{\bf Proof of Proposition \ref{main1}.}
The second equation in \eqref{EP3} follows directly from \eqref{gam}.
Now~rep\-lacing $\Gamma_k$ in the right hand side of \eqref{EP2} by $\Omega_{k-1}^T \Gamma_{k-1} \Omega_{k-1}$
and using the identity $\Omega_{k-1}\Omega_{k-1}^T ={\bf I}$, we get
\begin{gather*}
M_k + \frac {m\rho^2} 2 [ \Omega_k  \Gamma_k+ \Gamma_k  \Omega_k  ]
- \frac {m\rho^2} 2 [ \Omega_k^T  \Gamma_k+ \Gamma_k  \Omega_k^T  ]  = \Omega_{k-1}^T M_{k-1} \Omega_{k-1}
\\ \phantom{M_k}
{} + \frac {m\rho^2} 2 [\Gamma_{k-1}\Omega_{k-1} + \Omega_{k-1}^T \Gamma_{k-1} \Omega^2_{k-1} ]-
\frac {m\rho^2} 2 [ (\Omega_{k-1}^T)^2 \Gamma_{k-1} \Omega_{k-1}+  \Omega_{k-1}^T  \Gamma_{k-1}] ,
\end{gather*}
which, in view of \eqref{dis_K}, is equivalent to the f\/irst equation in \eqref{EP3}. \hfill $\blacksquare$

\begin{remark}
The form of equations \eqref{EP3}, \eqref{dis_K} is reminiscent to that of the discrete Euler
top on $SO(n)$ f\/irst described in \cite{Ves},
\begin{gather}
\label{DT}
M_k= \Omega_{k-1}^T {M}_{k-1} \Omega_{k-1}, \qquad M_k=  \Omega_k J - J\Omega_k^T ,
\end{gather}
however, the discrete momentum $\mathcal K$ in \eqref{dis_K} depends not only on $\Omega$, but also on $\gamma$.
Note that, in view of Proposition \ref{main1}, the subsequent momenta ${\mathcal K}_{k-1}, {\mathcal K}_k$
admit curious intertwining relations
\begin{gather*}
{\mathcal K}_{k-1}  = \Omega_{k-1} \biggl(J+ \frac D2 (\Gamma_{k-1}+\Gamma_k) \biggr)
-\biggl(J+ \frac D2 (\Gamma_{k-1}+\Gamma_k) \biggr) \Omega_{k-1}^T,
\\
{\mathcal K}_k  = \biggl(J+ \frac D2 (\Gamma_{k-1}+\Gamma_k) \biggr) \Omega_{k-1}
-\Omega_{k-1}^T \biggl(J+ \frac D2(\Gamma_{k-1}+\Gamma_k) \biggr) \, ,
\end{gather*}
which are reminiscent to the following relation between the subsequent momenta of the discrete Euler top \eqref{DT}
(see \cite{MosVes})
\begin{gather*}
M_{k-1} = \Omega_{k-1} J - J\Omega_{k-1}^T, \qquad M_{k} = J \Omega_{k-1} - \Omega_{k-1}^T J .
\end{gather*}
\end{remark}

\section[Discrete Chaplygin sphere on $E(3)$ and its particular solutions]{Discrete
Chaplygin sphere on $\boldsymbol{E(3)}$ and its particular solutions}

In the case $n=3$ we can make use of the following parameterization of the body f\/inite rotations
\begin{gather}
\label{Rod}
\Omega = \begin{pmatrix}
q_{0}^{2}+q_{1}^{2}-q_{2}^{2}-q_{3}^{2} & 2(q_{1}q_{2} - q_{3}q_{0}) &
2(q_{1}q_{3} + q_{2}q_{0}) \\[1ex]
2(q_{1}q_{2}+ q_{3}q_{0}) & q_{0}^{2}+q_{2}^{2}-q_{1}^{2}-q_{3}^{2} &
2(q_{2}q_{3}- q_{0}q_{1}) \\[1ex]
2(q_{1}q_{3}- q_{2}q_{0}) & -2(q_{2}q_{3}-q_{0}q_{1}) &
q_{0}^{2}+q_{3}^{2}-q_{1}^{2}-q_{2}^{2}
\end{pmatrix} ,
\end{gather}
where $q_0$, $q_i$ are the Euler parameters subject to relations $q_0^2+ q_1^2+q_2^2+q_3^2=1$ and $q_0>0$.
The operator \eqref{Rod} describes a f\/inite rotation in ${\mathbb R}^3$
about the vector ${\bf q} = (q_1, q_2, q_3)^T$ by the angle $\theta$ such that $q_0=\cos (\theta/2)$ and
$|{\bf q}|=\sin(\theta/2)$ (see, e.g., \cite{Whitt}).

Let now $\vec M_k, \vec {\mathcal K}_k \in {\mathbb R}^3$ be the vector representations of
$M_k, {\mathcal K}_k \in so^*(3)$,
\begin{gather*}
\vec M  =(M_1,M_2, M_3)^T \equiv (M_{32}, M_{13},M_{21})^T,
\\
\vec {\mathcal K}  =( {\mathcal K}_1, {\mathcal K}_2, {\mathcal K}_3)^T \equiv
( {\mathcal K}_{32},  {\mathcal K}_{13}, {\mathcal K}_{21})^T .
\end{gather*}
Then, in view of \eqref{dis_K}, we get simple expressions
\begin{gather} \label{Mk}
\vec M = 2 \begin{pmatrix} (J_2+J_3) q_0 q_1 +(J_2-J_3) q_2q_3 \\[1ex]
                    (J_1+J_3) q_0 q_2 +(J_3 -J_1) q_1 q_3 \\[1ex]
                    (J_1+J_2) q_0 q_3 + (J_1-J_2) q_1q_2 \end{pmatrix},     \qquad
\vec {\mathcal K}= \vec M + 2 D q_0 ({\bf q}- \langle \gamma, {\bf q} \rangle \gamma ) \, .
\end{gather}
In the continuous limit, when the angle $\theta$ is small, $\theta= \varepsilon \omega$, $\varepsilon\ll 1$, we have
${\bf q}= \frac \varepsilon 2 \omega+O(\varepsilon^3)$, $q_0 =1-O(\varepsilon^2)$, and,
up to terms of order $\varepsilon$,
\begin{gather*}
\frac 1\varepsilon \vec {\mathcal K}= \Lambda \omega- D\langle \gamma, \omega \rangle \gamma , \qquad
\Lambda= \mbox{diag} (J_2+J_3+D, J_1+J_3+D, J_1+J_2+D ),
\end{gather*}
which coincides with the expression for the angular momentum vector $K$ in the classical Chaplygin problem.

As a result, in the vector form the map \eqref{EP3} reads
\begin{gather}
\label{Mom2}
\vec {\mathcal K}_k = \Omega^T_{k-1} \vec {\mathcal K}_{k-1}, \qquad \gamma_k = \Omega^T_{k-1}\gamma_{k-1} ,
\end{gather}
where  $\Omega_k$ is  recovered from ${\mathcal K}_k$, $\gamma_k$ by solving \eqref{Mk} with respect to $\bf q$ and
substituting the solution into \eqref{Rod}. One can show that
for real ${\mathcal K}$, $\gamma$, the equations \eqref{Mk} have at most 4 and at least 2~real solutions ${\bf q}$.

Like in most of other discrete systems, in order to choose one of the branches of the map~\eqref{EP3} one should restrict
to the case of suf\/f\/iciently small f\/inite rotations $\Omega_k$. In this case only one of the above real solutions
${\bf q}$ will be small and should be taken as the appropriate branch.

Like the continuous system \eqref{vectors},
apart from the geometric condition $\langle\gamma,\gamma\rangle=1$, the discrete system
 preserves two independent integrals
\begin{gather*}
\langle {\mathcal K} ,\gamma\rangle =h, \qquad \langle {\mathcal K} ,{\mathcal K} \rangle =n .
\end{gather*}
However, as simple numerical tests show, the energy integral \eqref{en_int} is not preserved.

{\bf The special case $\vec {\mathcal K}\parallel \gamma$.}
Like the continuous Chaplygin system, the map \eqref{Mom2} has the special case when
 $\vec {\mathcal K}_k=h\gamma_k$, $h=$const. In view of \eqref{Mk}, this implies the following
relation between $\gamma$ and ${\bf q}$
\begin{gather}
\label{vert}
2 \begin{pmatrix} ( {\widehat J}_2+{\widehat  J}_3 ) q_0 q_1 +({\widehat  J}_2-{\widehat  J}_3) q_2q_3 \\[1ex]
                    ({\widehat  J}_1+{\widehat  J}_3) q_0 q_2 +({\widehat  J}_3 -{\widehat  J}_1) q_1 q_3 \\[1ex]
                ({\widehat  J}_1+{\widehat  J}_2) q_0 q_3 + ({\widehat  J}_1-{\widehat  J}_2) q_1q_2 \end{pmatrix}
 = [h+ D\, q_0 \langle \gamma, {\bf q} \rangle] \gamma, \qquad \widehat  J =J +\frac D2 I\, .
\end{gather}

Then one obtains the reduced map ${\mathcal G}_h: \; S^2\mapsto S^2$, ${\mathcal G}_h ( \gamma_{k-1} )= \gamma_k$ evaluated
as follows:
\begin{enumerate}\itemsep=0pt
\item[1)] given $\gamma_{k-1}$, one recovers ${\bf q}$ as a solution of \eqref{vert},

\item[2)] one calculates $\Omega_k$ by \eqref{Rod} and $\gamma_k$ by the second equation in \eqref{Mom2}.
\end{enumerate}

\begin{proposition} \label{main2} For any branch of the map ${\mathcal G}_h$, it has the quadratic integral
\begin{gather} \label{quad1}
\langle \gamma, {\Lambda}^{-1} \gamma \rangle=l .
\end{gather}
\end{proposition}

\begin{proof} In view of \eqref{vert}, \eqref{Rod}, and the second equation in \eqref{Mom2}\footnote{To simplify notation, here and below we omit the discrete time index $k$ in the components of ${\bf q}$.},
\begin{gather}
\gamma_k = \frac 2{\sigma_k} \begin{pmatrix} ({\widehat J}_2+{\widehat  J}_3) q_0 q_1
+({\widehat  J}_2-{\widehat  J}_3) q_2q_3
\\[1ex]
({\widehat  J}_1+{\widehat  J}_3) q_0 q_2 +({\widehat  J}_3 -{\widehat  J}_1) q_1 q_3
\\[1ex]
({\widehat  J}_1+{\widehat  J}_2) q_0 q_3 + ({\widehat  J}_1-{\widehat  J}_2) q_1q_2 \end{pmatrix},
\qquad \sigma_k=[h+ D\langle \gamma_k, q_0{\bf q} \rangle],\nonumber
\\[2ex]
\gamma_{k+1} = - \Omega^T_{k}\gamma_{k} \equiv \frac 2{\sigma_k}
\begin{pmatrix} ({\widehat J}_2+{\widehat  J}_3) q_0 q_1 - ({\widehat  J}_2-{\widehat  J}_3) q_2q_3
\\[1ex]
({\widehat  J}_1+{\widehat  J}_3) q_0 q_2 -({\widehat  J}_3 -{\widehat  J}_1) q_1 q_3
\\[1ex]
({\widehat  J}_1+{\widehat  J}_2) q_0 q_3 - ({\widehat  J}_1-{\widehat  J}_2) q_1q_2 \end{pmatrix}.
\label{gamas}
\end{gather}
Substituting these expressions into \eqref{quad1}, we obtain $
\langle \gamma_k, { \Lambda}^{-1} \gamma_k \rangle = \langle
\gamma_{k+1}, {\Lambda}^{-1} \gamma_{k+1} \rangle$.
\end{proof}

As a result, in our special case the complex invariant variety of the map ${\mathcal G}_h$
is the spatial elliptic curve $\cal E$, the intersection
of the unit sphere $\langle\gamma,\gamma\rangle=1$ with the quadric given by \eqref{quad1}. The curve $\cal E$ is 4-fold
unramif\/ied covering of the planar elliptic curve
\begin{gather*}
{\mathcal E}_0 =\big\{ w^2= - (z-\Lambda_1^{-1})(z-\Lambda_2^{-1})(z-\Lambda_3^{-1})(z-l) \big\}
\end{gather*}
and the points of $\cal E$ are parameterized by the Jacobi elliptic functions associated
to ${\mathcal E}_0$. Assume for concreteness that
$\Lambda_1^{-1} < \Lambda_2^{-1} < l <\Lambda_3^{-1}$. Then the parameterization reads (see e.g., \cite{Whitt, Law})
\begin{gather}
\label{param}
\gamma_1= C_1 \,\mbox{cn} (u|k), \qquad \gamma_2=C_2\, \mbox{sn} (u|k), \qquad \gamma_3=C_3\, \mbox{dn}(u|k),
\\[1ex]
C_1= \sqrt{\frac {\Lambda_1 (1-\Lambda_3 l)}{ \Lambda_1-\Lambda_3 } }, \qquad
C_2= \sqrt{\frac {\Lambda_2 (1-\Lambda_3 l)}{\Lambda_2 -\Lambda_3 } }, \qquad
C_3= \sqrt{\frac {\Lambda_3 (1-\Lambda_1 l) }{ \Lambda_1-\Lambda_3 } },
\label{Cs}
\end{gather}
where $u$ is a complex phase parameter and $k$ is the modulus of ${\cal E}_0$ given by
\begin{gather*}
k^2= \frac {(\Lambda_1-\Lambda_2)(1-\Lambda_3 l)} {(\Lambda_2-\Lambda_3)(\Lambda_1 l-1) } .
\end{gather*}
Therefore, for a f\/ixed
$l$, the map ${\mathcal G}_h$ is reduced to one-dimensional map
\begin{gather}
\label{delta}
u_{k+1} =u_k + \Delta u_k (u_k, l)  ,
\end{gather}
with the increment $\Delta u_k$ is a function of $u_k$, $l$ to be determined below.

\begin{remark}
As seen from \eqref{gamas}, the structure of the map ${\mathcal G}_h$
is similar to that of the Veselov--Moser discretization \eqref{DT} of the Euler top on $so(3)$.
As shown in \cite{MosVes, Bob_Lorb}, the latter discretization
preserves the momentum and the energy integrals, whereas
the corresponding increment $\Delta u_k$
does not depend on the argument $u_k$ in the elliptic parameterization.
The same holds for another exact discretization of the top found in \cite{JNMP_Fed}.

In comparison with the Veselov--Moser discretization,
our map~${\mathcal G}_h$ involves the extra factor~$\sigma_k$, which is not constant on the orbits
of~${\mathcal G}_h$. This gives an indication that under our map~${\mathcal G}_h$ the increment $\Delta u_k$ depends
essentially on~$u_k$. Moreover, as ${\mathcal G}_h$ is multi-valued, the function $\Delta u_k (u_k, l)$ is expected to
be multi-valued as well.  This stays in contrast with the
solutions of the continuous equations \eqref{euler} obtained from the Chaplygin sphere system under the condition
$K\parallel \gamma$: the components of $\gamma$ are elliptic functions whose argument changes uniformly with time~$t$.
\end{remark}

In this connection the following natural problem of the complete integrability of the map ${\mathcal G}_h$ arises:
is there a (multi-valued) reparameterization $u=f(s,l)$, $s\in {\mathbb C}$ such that
\begin{gather*}
 u_k=f(s_k), \qquad u_{k+1}= f(s_k+\Delta s)
\end{gather*}
and the increment $\Delta s$ does not depend on $s_k$? Note that this property
is equivalent to that the one-dimensional map \eqref{delta} preserves the invariant measure $\mu(u) du$,
where the density $\mu(u)$ must satisfy the discrete Liouville equation
\begin{gather*}
 \frac {\partial u_{k+1}} {\partial u_{k}}= \frac{\mu(u_k)}{\mu(u_{k+1})}
\end{gather*}
and is related to $f(s)$ by the formula $df/ds=1/\mu$. Due to the existence of the integral \eqref{quad1},
this is also equivalent to that the map ${\mathcal G}_h$ itself preserves an invariant measure on $S^2$.

Since $u$ is the argument of an elliptic function, the restriction of the map \eqref{delta} onto the real axis
describes a dif\/feomorphism of a circle with the angular function $\Delta u_k (u_k, l)$.
Hence, the answer to the above integrability question requires studying the properties of the dif\/feomorphism and applying
known theorems in this f\/ield (see, e.g., \cite{Perez} and references therein).

{\bf Determination of $\Delta u_k (u_k,l)$.} For convenience, in the sequel we omit the
discrete time index~$k$ and denote ${\mathcal G}_h(\gamma)=\tilde\gamma$, $u+\Delta u =\tilde u$.
To determine the function $\Delta u (u,l)$, we shall use the following property.

\begin{proposition} \label{symm} The map ${\mathcal G}_h$ admits the following implicit form
that involves the sum and the difference of $\gamma$, $\tilde\gamma$ only:
\begin{gather}
\label{sp_dis}
  \tilde\gamma - \gamma = \frac \varkappa 4 \; ( \tilde\gamma + \gamma )
\times {\Lambda}^{-1} (\tilde \gamma + \gamma ),
\\
\varkappa = \frac{2 \sigma }{ 1+\sqrt{1-\dfrac 14
\langle \tilde \gamma + \gamma , {\Lambda}^{-2} (\tilde\gamma + \gamma )\rangle }  } ,\nonumber
\\
\sigma = \frac h {1- \dfrac{D}8 \langle \tilde\gamma +\gamma,
{\Lambda}^{-1} (\tilde\gamma +\gamma ) \rangle } .
\label{sigmas}
\end{gather}
\end{proposition}

\begin{proof} As follows from \eqref{gamas},
\begin{gather}
\label{g_q}
\tilde \gamma + \gamma= \frac {4q_0}{\sigma} \Lambda {\bf q}, \qquad
 \tilde  \gamma- \gamma = \frac 4{\sigma} \Lambda {\bf q} \times {\bf q}
\end{gather}
and, therefore,
\begin{gather*}
\tilde\gamma- \gamma = \frac{\sigma}{4q_0^2 } (\tilde\gamma + \gamma )
\times {\Lambda}^{-1} (\tilde\gamma + \gamma ).
\end{gather*}
Using \eqref{g_q}, it is easy to check that
\begin{gather*}
2 q_0^2 = 1+\sqrt{1-\frac 14
\langle {\Lambda}^{-1} (\tilde\gamma + \gamma ), {\Lambda}^{-1} (\tilde\gamma + \gamma )\rangle }
\end{gather*}
and that the factor $\sigma$ def\/ined in \eqref{gamas} is the solution of the equation
\begin{gather*}
\sigma = h+D \biggl\langle \gamma, \frac{\sigma}4 \Lambda^{-1} (\tilde\gamma + \gamma )\biggr\rangle.
\end{gather*}
Then, in view of relation
\begin{gather*}
 \langle \gamma, \Lambda^{-1} (\tilde\gamma + \gamma )\rangle = \frac 12 \langle \gamma + \tilde\gamma,
{\Lambda}^{-1} (\gamma +\tilde \gamma ) \rangle  ,
\end{gather*}
one obtains \eqref{sp_dis}, \eqref{sigmas}.
\end{proof}

Below we assume $h=1$ in \eqref{sigmas}, which can always be made by appropriate rescaling of
$\Lambda$ and $D$.

Implicit maps in the symmetric form \eqref{sp_dis} were studied in \cite{Bob_Lorb}
as special discretizations of the Euler
top on $so(3)$. The symmetry allows to determine the function $\Delta u (u,l)$ in \eqref{delta}. Following~\cite{Bob_Lorb}, we set
\begin{gather*}
\bar u= \frac{u+\tilde u}2, \qquad \delta u=\frac{\tilde u-u}2.
\end{gather*}
Then, due to addition formulas for the Jacobi elliptic functions (see e.g., \cite{Law}),
the parameterization \eqref{param}, \eqref{Cs} implies
\begin{gather}
\tilde\gamma_1 - \gamma_1 = \frac{2C_1}{\mathcal D}
{ \mbox{sn} (\bar u) \mbox{dn} (\bar u) \mbox{sn} (\delta u) \mbox{dn} (\delta u)} , \qquad
  \tilde\gamma_1 + \gamma_1 =\frac{2C_1}{\mathcal D}
\mbox{cn} (\bar u) \mbox{cn} (\delta u) , \nonumber \\
{\mathcal D}= 1-k^2 \mbox{sn} ^2 (\bar u)\, \mbox{sn}^2(\delta u)
\label{dd}
\end{gather}
and similar expressions for the other components of $\tilde\gamma
- \gamma$, $\tilde\gamma + \gamma$. Substituting them into any of
the equations \eqref{sp_dis}, one obtains (see also
\cite{Bob_Lorb})
\begin{gather*}
%\label{rel1}
\frac {2{\mathcal D}}\varkappa
= V \frac {\mbox{cn} (\delta u)\mbox{dn}(\delta u)}{\mbox{sn} (\delta u) },
\qquad V^2 = \frac{(\Lambda_2- \Lambda_3)(\Lambda_1 l -1)}{ \Lambda_1 \Lambda_2 \Lambda_3 }
\end{gather*}
or, in view of \eqref{sigmas},
\begin{gather}
\biggl(1 -\frac d 2 {\Pi_1} \biggr) \sqrt{ {\mathcal D}^2-\Pi_2 {\mathcal D}^2}
=V \frac {\mbox{cn} (\delta u)\mbox{dn}(\delta u)}{\mbox{sn} (\delta u) }
-{\mathcal D}\biggl(1 -\frac d 2 {\Pi_1} \biggr) ,
\label{str}
\\
\Pi_1= \frac 14 \langle \tilde\gamma + \gamma, {\Lambda}^{-1} (\tilde\gamma + \gamma )\rangle,
\qquad
\Pi_2= \frac 14
\langle \tilde\gamma + \gamma , {\Lambda}^{-2} (\tilde\gamma + \gamma )\rangle .\nonumber
\end{gather}
Then, making squares of both parts of \eqref{str} and taking into account the expressions
\begin{gather*}
\Pi_1 =
\frac 1{\mathcal D} \biggl(
\frac {C_1^2}{\Lambda_1}+\frac{C_3^2}{\Lambda_3}-\frac{C_2^2}{\Lambda_2} \mbox{sn}^2(\delta u) \biggr)
\equiv \frac 1{\mathcal D} \biggl(
l- \mu \mbox{sn}^2(\delta u) \biggr), \qquad \mu = \frac{1-\Lambda_3 l}{\Lambda_2-\Lambda_3}  ,
\\
\Pi_2  = \frac 1{{\mathcal D}^2}
 \bigg( \frac{C_1^2}{\Lambda_1^2} \mbox{cn}^2 (\bar u) \mbox{cn}^2 (\delta u)
+ \frac{C_2^2}{\Lambda_2^2} \mbox{sn}^2 (\bar u) \mbox{cn}^2(\delta u) \mbox{dn}^2(\delta u)+
 \frac{C_3^2}{\Lambda_3^2} \mbox{dn}^2 (\bar u) \mbox{dn}^2(\delta u)\bigg)
 \\ \phantom{\Pi_2}
{} \equiv \frac 1{{\mathcal D}^2} \bigg(\frac {G}{ \mbox{sn}^2(\delta u)} {\mathcal D}-
V^2 \frac { \mbox{cn}^2 (\delta u)\mbox{dn}^2(\delta u)}{\mbox{sn}^2 (\delta u) } \bigg)  ,
\end{gather*}
where
\begin{gather*}
G  = \frac{C_1^2}{k^2 \Lambda_1^2} \mbox{cn}^2(\delta u)
-\frac{C_2^2}{\Lambda_2^2} \mbox{cn}^2 (\delta u) \mbox{dn}^2 (\delta u)
+\frac{C_3^2}{\Lambda_3^2} \mbox{dn}^2 (\delta u)
\equiv V^2 (1 + \alpha\, \mbox{sn}^2 (\delta u)- \beta\, \mbox{sn}^4 (\delta u)),
\\
\alpha = \frac {2\Lambda_1 \Lambda_3
-(\Lambda_1+\Lambda_3-\Lambda_2)}{(\Lambda_2-\Lambda_3)(\Lambda_1 l-1) }, \qquad
\beta= \frac{\Lambda_1 \Lambda_3}{(\Lambda_2-\Lambda_3)^2}\frac{1-\Lambda_3 l}{\Lambda_1 l-1} ,
\end{gather*}
after simplif\/ications we obtain the following relation between the Jacobi elliptic functions of $\delta u$ and
the function sn($\bar u$) (contained in $\cal D$ only)
\begin{gather}
- \frac {d^2}{4} \frac { (l-\mu\, \mbox{sn}^2(\delta u) )^2}{{\mathcal D}^2}
( G{\mathcal D}-V^2 \mbox{cn}^2 (\delta u)\mbox{dn}^2 (\delta u)) \qquad \qquad \nonumber
\\ \qquad
 {}+d \frac {l\!-\!\mu \mbox{sn}^2(\delta u)}{\cal D}
( (G {\mathcal D} \!-\!V^2 \mbox{cn}^2 (\delta u)\mbox{dn}^2 (\delta u))
\!-\! Vw(\delta u) {\mathcal D} ) \!+\! {\mathcal D} (2V w(\delta u)\!-\!G)\!=\!0 ,
\label{comp}
\\
 w(\delta u)= \mbox{sn}(\delta u) \mbox{cn}(\delta u)\mbox{dn} (\delta u)  . \nonumber
\end{gather}
Note that in the case $d=0$, when the map ${\mathcal G}_h$ reduces to the discretization of the Euler top,
the terms with $\cal D$ in \eqref{comp} can be eliminated and the increment $\delta u$ becomes
independent of $\bar u$, as expected and was shown explicitly in \cite{Bob_Lorb}.

Now, expressing sn$(\bar u)\equiv$sn$(u+\delta u)$ in terms of elliptic functions of $u$ and $\delta u$,
and using relations cn$(u)^2=1-\mbox{sn}^2(u)$, dn$(u)^2=1-k^2 \mbox{sn}^2(u)$,  we can rewrite \eqref{dd} as
\begin{gather}
{\cal D}  = \frac 1 {[1-k^2 \mbox{sn} ^2 (u) \, \mbox{sn}^2(\delta u)  ]^2}
 ( k^2 \mbox{sn} ^2 (u) \mbox{sn} ^2 (\delta u)
(2-(k^2+1) \mbox{sn}^2 (\delta u) -k^2 \mbox{sn}^4 (\delta u) ) \nonumber
\\ \phantom{{\cal D}  =}
{} + 2k^2 \mbox{sn}^2 (\delta u) w(\delta u) W(u)
+k^2 \mbox{sn}^4 (\delta u) [1-k^2 \mbox{sn}^2(u)  ]-1 ) , \nonumber
\\
W(u) = \mbox{sn}(u) \mbox{cn}(u) \mbox{dn}(u)\, , \qquad
w(\delta u)= \mbox{sn}(\delta u) \mbox{cn}(\delta u)\mbox{dn} (\delta u)  . \label{calD}
\end{gather}
Substituting this expression into \eqref{comp}, one obtains a
rather complicated equation for $\mbox{sn}^2(u)$,
$\mbox{sn}^2(\delta u)$, $w(\delta u)$ that implicitly describes
the dependence the function $\Delta u_k (u_k,l)$.

One can make this equation algebraic by introducing the variables $p=\wp (\delta u)$, $q=\wp '(\delta u)$, which are
coordinates on the curve $\cal E$ represented in the canonical Weierstrass form
\begin{gather*}
 q^2 = 4 (p-e_1)(p-e_2)(p-e_3), \quad
e_1= \frac 13 (2-k^2), \quad e_2= \frac 13 (2k^2-1), \quad e_3 = -\frac 13 (1+k^2) .
\end{gather*}
Under the rational parameterization
\begin{gather*}
\mbox{sn}^2 (\delta u)= \frac 1{p-e_3},
\qquad  \mbox{cn}^2 (\delta u)= \frac {p-e_1}{p-e_3},
\qquad \mbox{dn}^2 (\delta u)= \frac {p-e_2}{p-e_3}, \qquad
w(\delta_u) = \frac 12 \frac q{(p-e_3)^3}
\end{gather*}
equation \eqref{comp} together with \eqref{calD} def\/ines a plane algebraic curve whose
intersections $P^{(j)}$ with ${\mathcal E}_0$ give the values of $\delta u$ via the integral
\begin{gather*}
\delta u^{(j)} = \int_\infty^{P^{(j)}} \frac {dp} {2\sqrt{(p-e_1)(p-e_2)(p-e_3)}} \, .
\end{gather*}

\section{Conclusive remarks}
In given paper we constructed a discretization of the Chaplygin
sphere problem on $E(n)$ which preserves all the momenta integrals
and, in the particular case of the motion, the energy. This case
is reduced to the multi-valued map ${\mathcal G}_h$  on $so^*(3)$
which is reminiscent to the Veselov--Moser discretization of the
Euler top and is given by an addition law on elliptic curves.
However, in contrast to known integrable maps of this type, the
increment depends not only on the energy constant, but also on the
point on the curve. According to \cite{Bob_Lorb}, this implies
that ${\mathcal G}_h$ does not preserve the standard Lie--Poisson
structure on $so^*(3)$, which is quite expected due to the
nonholonomic origin of the map.

The integrability of the map ${\mathcal G}_h$ (which is interpreted as the
possibility of writing its explicit discrete time solution) was shown to be
equivalent to preservation of an invariant measure on the Lie algebra. Note that the latter property
is often related to integrability in quadratures in the continuous case
(recall the Jacobi last multiplier theorem).

In this connection it would be interesting to develop integrability criteria for discrete systems preserving
an invariant measure, in particular to formulate a discrete version of the Jacobi last multiplier theorem.

Although the energy \eqref{en_int} of the Chaplygin sphere
expressed in terms of the momentum $\mathcal K$ is not preserved
in the discrete setting in general, this does not exclude the
existence of another integral that transforms to the energy in the
continuous limit.

Lastly, as shown in \cite{Schneider}, the classical Chaplygin sphere problem is a natural example of
a~nonholonomic
LR system on the direct product $SO(3) \times {\mathbb R}^3$ (the kinetic energy is left-invariant, while the
constraint distribution is right-invariant with respect to the group action). One can show that the discrete
Lagrangian \eqref{disc_L} and the discrete constraint subvariety given by \eqref{dc} possess the same properties, i.e.,
our discretization of the Chaplygin sphere is a discrete LR system on $SO(n) \times {\mathbb R}^n$.

We then believe that it is worth studying properties of generic
discrete LR systems and of their reductions, such as preservation of momenta, energy, and an invariant measure.

\subsection*{Acknowledgments} I am grateful to the anonymous referees whose remarks helped to improve the text.
The research was partially
supported by Spanish Ministry of Science and Technology grant BFM 2003-09504-C02-02.

\pdfbookmark[1]{References}{ref}
\LastPageEnding

\end{document}